\documentclass{aa}

\usepackage{graphicx,natbib}
\usepackage{txfonts}
\usepackage{amsmath,amssymb}
\bibpunct{(}{)}{;}{a}{}{,}

\begin{document}

\title{Seed particle formation for silicate dust condensation by SiO nucleation.}


\author{
 Hans-Peter Gail\inst{1}
 \and Steffen Wetzel\inst{2}
 \and Annemarie Pucci\inst{2}
 \and Akemi Tamanai\inst{2}
}

\institute{
Universit\"at Heidelberg, Zentrum f\"ur Astronomie, 
           Institut f\"ur Theoretische Astrophysik,
           Albert-Ueberle-Str. 2,
           69120 Heidelberg, Germany 
\and
Universit\"at Heidelberg, Kirchhoff-Institut f\"ur Physik,
           Im Neuenheimer Feld 227,
           69120 Heidelberg, Germany 
  }

\offprints{\tt gail@ita.uni-heidelberg.de}

\date{Received date ; accepted date}

\abstract{Dust formation in stellar outflows is initiated by the formation of some seed particles that form the growth centres for macroscopic dust grains. The nature of the seed particles for silicate dust in stellar outflows with an oxygen rich element mixture is still an open question. Clustering of the abundant SiO molecules has several times been discussed as a possible mechanism and investigated both theoretically and by laboratory experiments. The initial results seemed to indicate, however, that condensation temperatures obtained by model calculations based on this mechanism are significant lower than what is really observed, which renders SiO nucleation unlikely.
}
{This negative result strongly rests on experimental data on vapour pressure of SiO. The case for SiO nucleation may be not as {\bf bad} as it previously seemed and needs to be re-discussed because new determinations of vapour pressure of SiO molecules over solid SiO have shown the older data on SiO vapour pressure to be seriously in error. Here we aim to check again in the light of improved new data the possibility that SiO nucleation triggers the cosmic silicate dust formation.
}
{First we present results of our measurements of vapour pressure of solid SiO. Second, we use the improved vapour pressure data to re-calibrate existing experimental data on SiO nucleation from the literature. Third, we use the re-calibrated data on SiO nucleation in a simple model program for dust-driven winds to determine the condensation temperature of silicate in stellar outflows from AGB stars.}
{Our measurements extend the temperature range of measurements for the vapour pressure to lower temperatures and pressures than before. This improves the reliability of the required extrapolation from the temperature range where laboratory data can be obtained to the temperature range where circumstellar dust condensation is observed. We determine an analytical fit for the nucleation rate of SiO from re-calibrated literature data and show that onset of nucleation under circumstellar conditions commences at higher temperature than was previously found.
This brings calculated condensation temperatures of silicate dust much closer to observed condensation temperatures derived from analysis of infrared spectra from dust enshrouded M stars. Calculated condensation temperatures are still by about 100 K lower than observed ones, but this may be due to the greenhouse effect of silicate dust temperatures which is not considered in our model calculation.}
{The assumption that the onset of dust formation in late-type stars with oxygen rich-rich element mixtures is triggered by cluster formation of SiO is compatible with dust condensation temperatures derived from IR observations.}


\maketitle

\titlerunning{SiO condensation in space}

\section{Introduction}

Outflow of matter in stellar winds of highly evolved stars is observed to be accompanied by copious dust formation. As long as the element mixture in the outflow remains oxygen-rich the dominating dust components are magnesium-iron-silicates \citep[see][ for a review on observations]{Mol10}. Since the outflowing gas from the star is free of condensation seeds, the first step in the sequence of events ultimately ending with the growth of silicate dust grains in the cooling outflow has to be the formation of seed particles. Because SiO is one of the most abundant of all the gas-phase species of refractory elements that could be involved in the process, it has several times been discussed if it is cluster formation by this species that stands at the base of all the processes \citep{Don81, Nut82, Nut83, Gai86, Gai98, Gai98b, Ali05, Nut06, Reb06, Paq11, Gou12} that ultimately leads to the condensation of silicate dust particles. 

The reason why formation of large complexes of SiO molecules or of solid SiO is thought to be an important intermediate step in the formation of silicate dust is as follows: There exist no molecular species with composition corresponding to the chemical formula of silicate compounds as free gas-phase species (at least not in detectable amounts). It is necessary, therefore, that the initial stages of the condensation of silicates involve some different kind of material. It has been thought since the early 1980s that condensation of SiO is the most obvious very initial step of silicate formation \citep[e.g.][]{Nut82,Gai86} because of the high abundance of SiO molecules in the gas phase in oxygen-rich environments, and in particular because there exists a rather stable condensed phase with chemical composition SiO. It is assumed that initially silicon monoxide particles form which then serve as seed particles for growth of silicates. There exists even the possibility that in a number of cases the silicon monoxide seed particles finally grow to a dust species of their own \citep{Wet12a}.

Some dedicated laboratory experiments have been performed \citep{Nut82} to study the nucleation and condensation of SiO. It turned out that the
results suggest an onset of condensation at a much too low temperature to be
compatible with observed condensation temperatures in circumstellar dust 
shells. Theoretical calculations \citep{Gai86} based on the results of
\citet{Nut82} show that nucleation  of SiO should not commence before
temperature has dropped below $\approx 600$\,K, while observed condensation
temperatures in circumstellar dust shells are typically of the order of
900~K or sometimes even higher. Alternatives to seed formation by SiO are nucleation of oxides of, e.g., Al or  Ti \citep{Gai98,Gai98b,Jeo98}. Such mechanisms, however, {\bf raise} serious problems. Because of low elemental abundances of Al or Ti, seed particle formation at mass-loss rates as low as $10^{-7}\,\rm M_{\sun}\,a^{-1}$ would occur at a {\bf prohibitively} low rate, while it is observed that even at such low mass-loss rates at least some dust can be formed. Hence, proposed alternatives to SiO nucleation were not successful. 

Therefore it seems worthwhile to re-investigate the case of SiO nucleation and
to look for possibilities to circumvent the difficulties with this mechanism that in the past lead to a rejection of the SiO-nucleation hypothesis. 

\citet{Nut06} and \citet{Fer08} addressed the problem of the vapour pressure of SiO and found this to be considerably lower than previous measurements \citep{Sch60} suggested. This would allow condensation of SiO to occur at higher temperatures than previously thought. They argued on this basis \citep{Nut06} that SiO could be the first species to nucleate in the outflow from a star with oxygen-rich element mixture if additionally effects of non-TE level population of SiO molecules \citep{Nut81} are accounted for. But even then condensation temperatures do not fit observations very well.

We also performed a new determination of the vapour pressure of SiO \citep{Wet12} and use the result to re-evaluate the old experimental results of \citet{Nut82} on SiO nucleation using our new data for SiO vapour pressure. We will show that the experimentally determined nucleation rates after correcting for the former strongly overestimated vapour pressure of SiO are significantly higher than those derived by using the old vapour pressure data of \citet{Sch60}. Extrapolated to circumstellar conditions, these corrected experimental data predict much higher condensation temperatures than previously.

Additionally we argue, that the previous model calculations of circumstellar dust condensation have not properly taken into account, that the ``dust condensation temperature'' derived from analysis of infrared spectra is the temperature of lattice vibrations of the dust particles in the region of onset of massive dust growth in the shell, while the nucleation process is governed by the gas temperature. We show that both temperatures may be significantly different.
Both effects cooperate to remove the hitherto suspected strong discrepancy between calculated nucleation temperatures and observed dust temperatures. This makes, again, clustering of SiO the hottest candidate for the production of seed particles for silicate dust condensation.

Another way to tackle the problem is to study the chemistry of the gas-phase and the kinetics of cluster formation on the basis of reaction kinetics \citep[cf.][]{Reb06,Gou12}. This is not considered here but will be the subject of a forthcoming paper

\begin{figure}[t]
\centerline{
\includegraphics[]{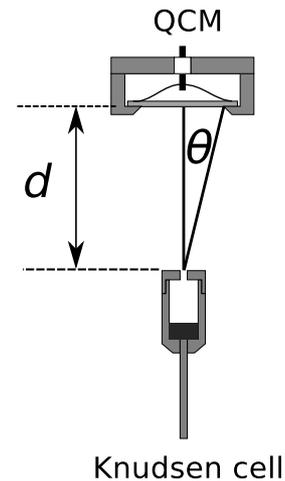}
}

\caption{
Schematic drawing of the experimental setup for the Knudsen method. The quartz crystal microbalance (QCM) is placed at a distance $d$ from the Knudsen cell and measures the mass rate of the condensing particles during effusion from the cell. 
}
\label{FigExpSet}
\end{figure}

\section{Vapour pressure of SiO}

Experimentally obtained vapour pressure data of SiO are discussed in this section, including a brief description of our experimental setup which is based on the Knudsen method \citep{Knudsen}. Fundamentally, the mass-loss rate and vapour pressure are measured by a quartz crystal micro balance and the Knudsen cell is heated by electron bombardment with an electron beam evaporator. All experimental results as well as detailed information regarding the experimental conditions and the samples are defined in \citet{Wet12}.

\subsection{Knudsen method}

The vapour pressure can be determined by measuring the mass-loss rate of a Knudsen cell, $({\mathrm{d}m}/{\mathrm{d}t})_\mathrm{K}$, at a well-defined temperature $T$. The Knudsen cell is a small container with a small orifice, $A_\mathrm{O}$, that is partially filled with an experimental sample. The vapour pressure inside the cell and the mass-loss rate are related by
\begin{equation}\label{eq:kinetic}
{P_\mathrm{K}}\left(T\right) = \frac{1}{A_\mathrm{O} W_\mathrm{O}} \left(\frac{\mathrm{d}m}{\mathrm{d}t}\right)_\mathrm{K} \sqrt{\frac{2 \pi R T}{M}},
\end{equation}
where $W_\mathrm{O}$ is the Clausing factor of the orifice, R is the ideal gas constant and M is the molecular mass \citep{Margrave}. The Clausing factor is the probability for a particle to pass the orifice and can be calculated for cylindrical tubes from its geometry \citep{Clausing32}. The equilibrium vapour pressure can be expressed by the Whitman- Motzfeld equation  
\begin{equation}\label{eq:whitmann}
 P\left(T\right) = \left[1 + \frac{A_\mathrm{O} W_\mathrm{O}}{A} \left( \frac{1}{\alpha\left(T\right)} + \frac{1}{W_\mathrm{Cell}} -2 \right)\right] P_\mathrm{K}, 
\end{equation}
where $\alpha\left(T\right)$ is the vaporisation coefficient, $A$ is the cross-sectional area of the cell, and $W_\mathrm{Cell}$ is the Clausing factor of the cell \citep{Whitman52, Motzfeldt55}. Both unknown quantities in Eq~(\ref{eq:whitmann}), which are the equilibrium vapour pressure and the vaporisation coefficient, can be derived from two measurements with different orifice sizes. 

\begin{figure}[t]

\centerline{
\includegraphics[width=0.9\hsize]{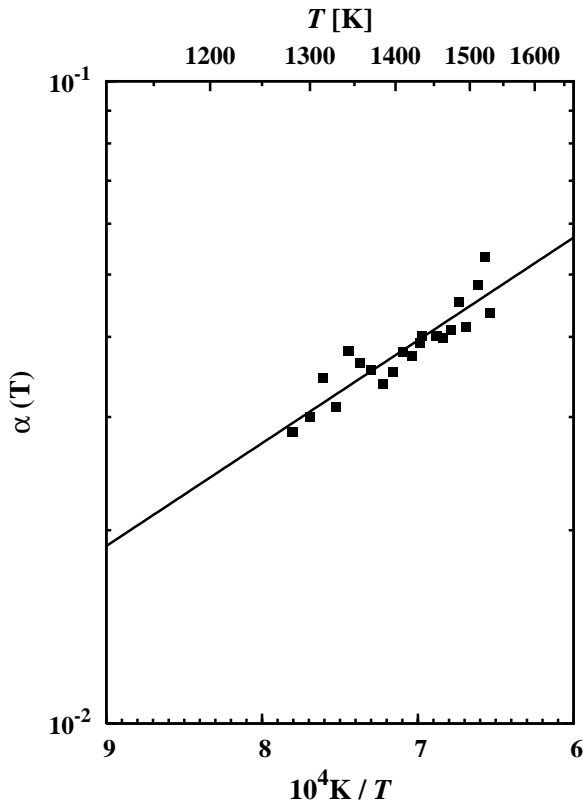}
}

\caption{Vaporisation coefficient of solid SiO derived from the experimental measurements of the mass-loss rate obtained for two orifice diameters of the Knudsen cell. The full line is the approximation according to Eq. (\ref{eq:fitalpha})}

\label{vapco}
\end{figure}

\subsection{Experimental setup}

Here we briefly describe our experimental setup \citep[for more details see][]{Wet12}. A basic schematic drawing of the experimental setup is shown in Fig.~\ref{FigExpSet}. The quartz crystal micro-balance, which is placed at a distance $d$ from the Knudsen cell, measures the mass changes due to condensing particles during the molecular effusion from the Knudsen cell. The angular dependence of the beam profile can be calculated from the equations given in \citet{Clausing30, Dayton56} which directly relate the signal detected from the quartz crystal micro-balance to the mass-loss of the Knudsen cell. The temperature is determined by a temperature calibration that gives a relation between the temperature in the cell and the heating power controlled during the measurements. The reliability and performance quality of this experimental approach has been verified by a precise reproduction of the vapour pressure of copper \citep{Wet12}. 

\subsection{Experimental Results}

Two orifices with different diameter (0.50~mm and 0.79~mm) of the Knudsen cell were employed in order to determine the vapour pressure as well as the vaporisation coefficient of SiO. 

Fig.~\ref{vapco} shows the vaporisation coefficient of SiO derived from the experimental outcomes as a function of temperature. The results can be approximated by an Arrhenius law
\begin{equation}
 \alpha\left(T\right)= \alpha_0\,{\rm e}^ {-T_0/T} \label{eq:fitalpha}
\end{equation}
with
$$
\alpha_0=0.521\pm0.156\,,\quad T_0=3\,686\pm428
$$
for the temperature range between 1275~K and 1525~K. This shows that the kinetics of vaporisation involves a modest activation energy barrier. 

\begin{figure}[t]

\includegraphics[width=.95\hsize]{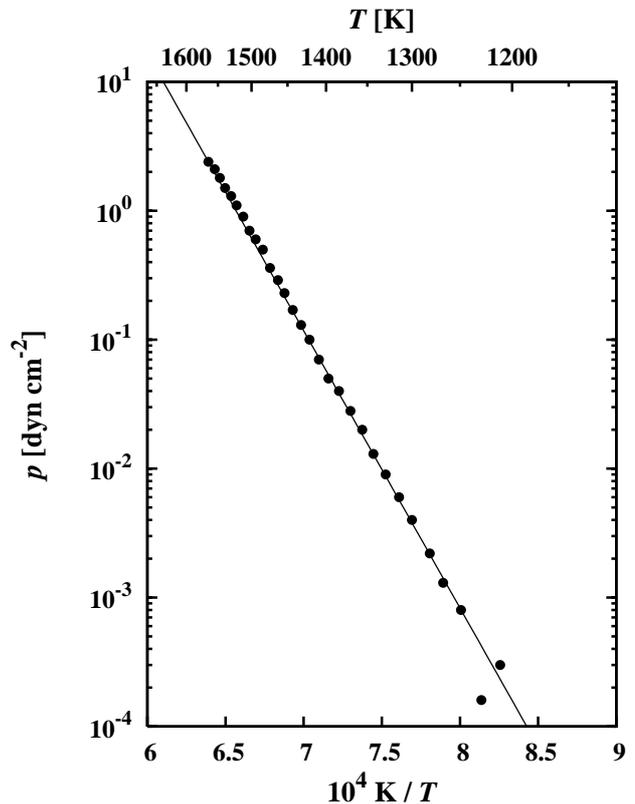}

\caption{Equilibrium vapour pressure of solid SiO derived from the experimental measurements of the mass-loss rate obtained for two orifice diameters of the Knudsen cell (filled circles). The full line is the analytic approximation according to Eq. (\ref{VapPrSiO}).
}
\label{VapPr}
\end{figure}

The equilibrium vapour pressure of SiO over solid silicon monoxide was calculated by using the vaporisation coefficient described by Eq~(\ref{eq:fitalpha}). The equilibrium vapour pressure as a function of inverse temperature is plotted in Fig.~\ref{VapPr}. From a best fit to the experimental data points, the vapour pressure of SiO molecule over solid silicon monoxide  in units of dyn\,cm$^{-2}$ is approximated by
\begin{equation}
P=\exp\left(-{T_v\over T}+S_v\right)\,.
\label{VapPrSiO}
\end{equation}
with
$$
T_v=49\,520\pm1\,400\,{\rm K}\,,\quad S_v=32.52\pm0.97\,.
$$

\section{Stability limits}

\begin{figure}

\includegraphics[width=\hsize]{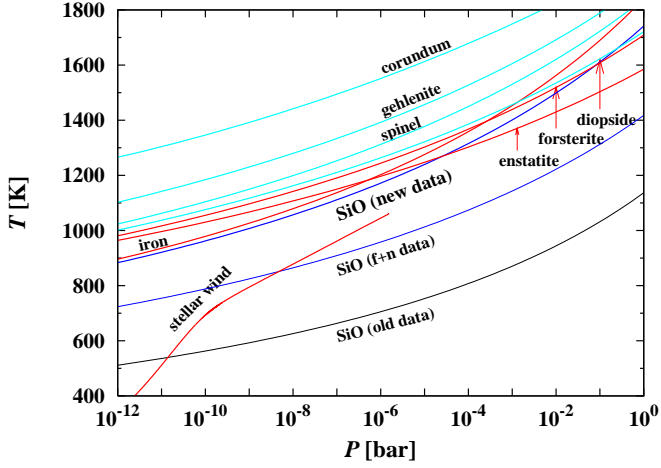}

\caption{Stability limit of solid SiO for the new data from our laboratory measurements, for data from \citet{Fer08} and for the old data of \cite{Sch60}. Also shown are stability limits for the most abundant refractory compounds formed from an oxygen-rich element mixture with cosmic standard element abundances. The dashed line corresponds to the $p$-$T$-trajectory of a dust-driven stellar wind.}

\label{FigLimCosm}
\end{figure}

Dust is formed in space in a variety of environments either in a continuous 
fashion like in stellar winds or brown dwarf atmospheres, or as short events
in stellar ejecta following explosive events like in supernovae or in novae.
The elemental composition of dust forming matter may also be very diverse in
different kinds of objects. Here we consider dust formation from oxygen rich
material with standard cosmic element mixture \citep[cf., e.g.,][]{Lod09}. The
most abundant dust materials that are expected to form from the most abundant 
refractory elements in such an element mixture, i.e., from Si, Mg, and Fe, are magnesium-rich silicates and metallic iron \citep[cf., e.g.,][]{Gai03}, and 
from the somewhat less abundant but even more refractory elements Al and Ca 
the minerals corundum (or hibonite), spinel, gehlenite, and diopside. The upper
stability limits or the field of stability in the $p$-$T$-plane of these
compounds are shown in Fig.~\ref{FigLimCosm}. They are calculated for chemical
equilibrium states between gas phase and condensed phases.

As a new feature we show the upper stability limit of SiO according to our new measurements of the vapour pressure of SiO molecules over solid SiO. This curve in the $p$-$T$-plane is constructed from the relation
\begin{equation}
p\,{\epsilon_{\rm Si}\over2+\epsilon_{\rm He}}=p_{\rm vap}(T)
\end{equation}
Here $p$ is the gas pressure and $p_{\rm vap}$ the vapour pressure of SiO molecules over solid SiO according to Eq.~(\ref{VapPrSiO}), and $\epsilon_{\rm Si}$ and $\epsilon_{\rm He}$ are the element abundances of Si and He with respect to H. The left hand side corresponds to the pressure of SiO molecules in the gas phase in an element mixture with cosmic composition. Here we made use of the fact that in the pressure and temperature regime of interest H is completely associated to H$_2$ and almost all Si is bound in SiO.%
\footnote{%
In an oxygen rich element mixture and at the temperatures of interest most of the Si is bound in SiO, but some of it is bound in SiS. This, as well as all other Si-bearing compounds, is neglected.}%
  
For comparison we also show in Fig.~\ref{FigLimCosm} the stability limits for solid SiO resulting from the vapour pressure data reported by \citet{Fer08}, i.e., calculated from Eq.~(\ref{VapPrSiO}), but with coefficients
\begin{equation}
T_v=40\,850\pm1\,270\,K\,,\quad S_v=32.93\pm0.90 
\end{equation}
(in units dyn\,cm$^{-2}$) and that resulting from the old vapour pressure data for solid SiO of \citet{Sch60}
\begin{equation}
p_{\mathrm{vap}}=\exp\left(26.66-{25670\over T}\right)
\label{PvapSiOSchick}
\end{equation}
(in units dyn\,cm$^{-2}$). From Fig.~\ref{FigLimCosm} it is immediately obvious that the condensation temperatures following from the data of \citet{Sch60} are {\bf unrealistically} low.

For our new data as well as for the recent data of \citet{Fer08} the stability limit of solid SiO is still at lower temperature than the upper stability limits of the more refractory silicates and the Ca-Al-compounds, but with the strongly reduced vapour pressure determined by our measurements and that of \citet{Fer08} it is not as unfavourably low for condensation under circumstellar conditions as it was suggested by the old data of \cite{Sch60}. The strong upward shift of the stability limit according to the new data makes SiO again a candidate for the seed particle formation in M stars. First steps to investigate this have already performed by \citet{Nut06} and \citet{Paq12}. Here we follow a different approach. 
 
\section{Application to SiO nucleation} 
\label{SectNuc}

The details of the reaction mechanisms standing behind the nucleation of SiO are not yet known, but some experimental results have been obtained by \citet{Nut82} which give insight into the conditions under which SiO nucleation occurs. It seems worthwhile to look how these results change if corrected for the much reduced vapour pressure of SiO.

\subsection{Re-evaluation of the Nuth \& Donn data}

In their laboratory experiment \citet{Nut82} evaporated solid SiO from a 
crucible into a hydrogen atmosphere in order to prepare a gas mixture that 
contains only SiO and H$_2$ molecules. In this case, any observed condensation
necessarily starts with cluster formation from SiO molecules.%
\footnote{For this reason we do not consider the results of \citet{Nut83} for the more complex Si-O-Mg system.}

They determined the temperature and the partial pressure of SiO at which an
onset of smoke formation can be observed. This was quantified by taking the 
instant, at which the extinction of a light beam with wavelength $\lambda=200\,
\rm nm$ traversing the condensation zone reached a level of 1\%, as the instant where avalanche nucleation commences. Unfortunately the rate of nucleation $J$ at the onset of detectable smoke formation could only crudely be estimated. It
was argued that the value of $J$ at this instant is between $10^8$ and 
$10^{10}$ particles cm$^{-3}$\,s$^{-1}$. 

The results were presented as a set of values of temperature $T$ and
supersaturation ratio
\begin{equation}
S={p_1\over p_{\mathrm{vap}}}
\end{equation}
with $p_1$ being the partial pressure of SiO in the condensation zone and
$p_{\mathrm{vap}}$ the vapour pressure of SiO molecules over solid SiO at
temperature $T$.
The vapour pressure was calculated by \cite{Nut82} from the formula given 
by \citet{Sch60}, Eq.~(\ref{PvapSiOSchick}). The partial pressure $p_1$ or particle densities $n_1=p_1/kT$ of SiO molecules in the gas phase in the condensation zone was calculated (i) from the vapour pressure of SiO in the crucible at the temperature of the crucible and (ii) from the dilution as the vapour diffused away from the crucible to the condensation zone. The pressure in the condensation zone was estimated to be 1/10-th of the vapour pressure at the crucible. 

\begin{figure}

\includegraphics[width=\hsize]{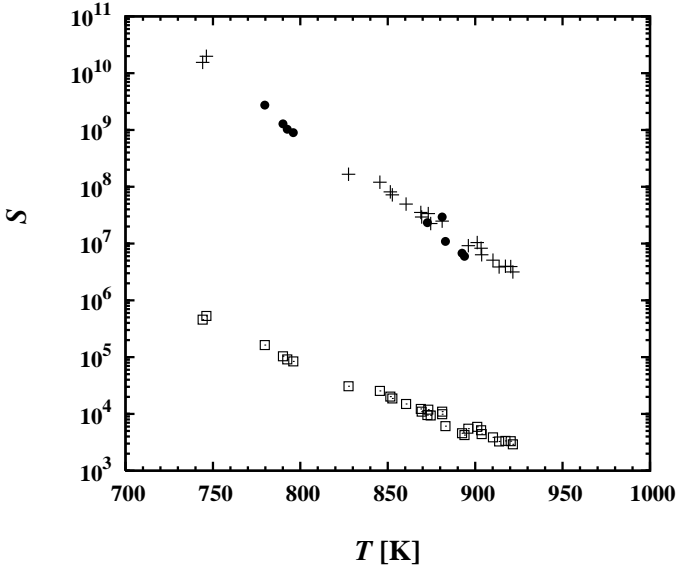}

\caption{Supersaturation at onset of avalanche nucleation from the experiment of
\citet{Nut82}, re-evaluated with the new vapour pressure data for SiO. Black 
dots: Data for the nucleation experiments conducted at $p_{\mathrm{H_2}}=50$\,Torr. Crosses: Data for experiments conducted at $p_{\mathrm{H_2}}=35$\,Torr. Open sqares: Supersaturation based on old vapour pressure data, as given by \citet{Nut82}, for the same set of experiments.
}

\label{FigSupNuthDonn}
\end{figure}

For a new evaluation of the experimental results of \citet{Nut82} we read-off values for $S$ and $T$ from their Fig.~3. We can convert these data for $S$ at given $T$ to a new set of data based on the improved vapour pressure formula for SiO in the following way: First we calculate from the temperature $T$ by means of Eq.~(\ref{PvapSiOSchick}) a vapour pressure $p_{\mathrm{vap}}$ and from the value of $S$ the corresponding pressure $p_1$. Considering that this was assumed to equal 1/10-th of $p_{\mathrm{vap}}$ calculated with the crucible temperature $T_{\mathrm{c}}$ we reconstruct from Eq.~(\ref{PvapSiOSchick}) the temperature $T_{\mathrm{c}}$. Then we calculate with this temperature a new value for $p_1$ at the crucible by means of Eq. (\ref{VapPrSiO}) and take 1/10 of this as the corrected value for $p_1$ in the condensation zone. From the same equation we calculate a new value of $p_{\mathrm{vap}}(T)$ in the condensation zone and from this and the corrected value of $p_1$ a new value for $S$. This defines a new set of values for $S$ at given $T$ that correspond to the fixed value of $J_{\rm av}=10^{9\pm1}$ particles cm$^{-3}$\,s$^{-1}$ in the experiments of \citet{Nut82}, but now based on the improved vapour pressure of solid SiO. 

The results are shown in Fig.~\ref{FigSupNuthDonn} for the two series of 
measurements with a total pressure of $p_{\mathrm{H_2}}=35$\,Torr and 
$p_{\mathrm{H_2}}=50$\,Torr. This looks very much the same as the corresponding
Fig. 3 of \citet{Nut82}, except that the new values for $S$ are more than a factor of thousand higher than those calculated with the old vapour pressure formula (\ref{PvapSiOSchick}).

\subsection{Fit for the nucleation rate}

Next we follow the same procedure as \citet{Nut82} and check if the results
can be fitted with the formula resulting from Zeldovich-theory of homogeneous
nucleation \citep{Zel43} that gives the following result for the nucleation 
rate
\begin{equation}
J=\left(2\sigma\over\pi m\right)^{1\over2}V \alpha n_1^2\exp\left(-{16\pi\sigma^3 V^2
\over 3k^3 T^3(\ln S)^2}\right)\,.
\label{NucRatClassic}
\end{equation} 
Here $\sigma$ is the surface energy of the condensed phase, $V$ the volume of the SiO molecule in the solid, $A$ and $m$ the atomic weight and mass of the SiO molecule, respectively, $\alpha$ the sticking coefficient at the size of the critical cluster, and $n_1=p_1/kT$ the particle density of SiO molecules in the gas phase. 

\begin{figure}

\includegraphics[width=\hsize]{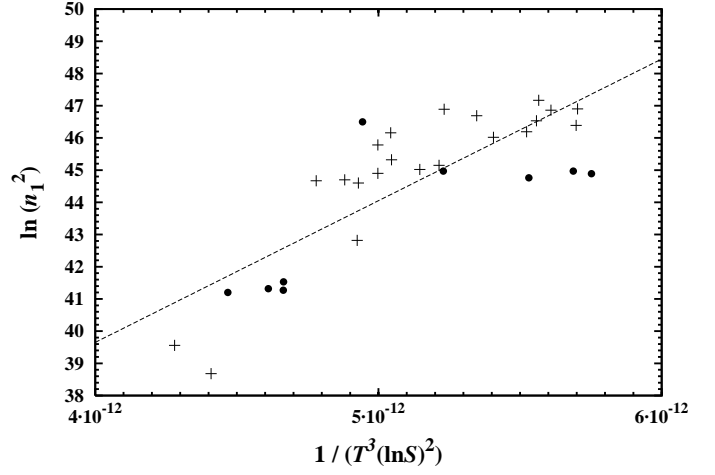}

\caption{Plot of $\ln n_1^2$ versus $[T^3(\ln S)^2]^{-1}$. The dashed line shows
the best fit of a linear relation, Eq. (\ref{FitLinSig}). Symbols as in 
Fig.~\ref{FigSupNuthDonn}.
}

\label{FigGamFit}
\end{figure}

Equation (\ref{NucRatClassic}) predicts that for a constant nucleation rate $J_{\rm av}$ as in the measurements of  \citet{Nut82} there should exist a linear relationship between $\ln n_1$ and $T^3\ln S$ of the form 
\begin{equation}
\ln n_1^2=a\,{1\over T^3(\ln S)^2}+b
\label{FitLinSig}
\end{equation}
with constants $a$ and $b$ given by
\begin{align}
a=&{16\pi\sigma^3 V^2\over 3k^3}\\
b=&\ln{J_{\rm av}\over \sqrt{2\sigma\over\pi m}V \alpha} 
\end{align}
This relationship rests on assumptions that may hold for macroscopic sized particles but are hardly to justify for molecular-sized particles. Nevertheless one can speculate that the dependency of the classical nucleation rate on the basic macroscopic parameters $T$, $S$, and $n_1$ is essentially correct, except that the constants in this equation cannot meaningful be derived within the frame of a macroscopic theory. This means that we assume that the nucleation rate $J$ is given by
\begin{equation}
J=n_1^2\exp\left(B-{a\over T^3(\ln S)^2}\right)
\end{equation}
with empirically to be determined constants $a$ and
\begin{equation}
B=\ln J_{\rm av}-b\,.
\end{equation}

One can check this hypothesis by fitting the experimental results of \citet{Nut82} for $n_1$ and $S$ for given $T$ by the linear relation (\ref{FitLinSig}). The result is shown in Fig.~\ref{FigGamFit} where $\ln n_1^2$ is plotted against $[T^3(\ln S)^2]^{-1}$ for the experimental data corrected for the new values for the vapour pressure; Fig.~\ref{FigGamFit} is completely analogous to Fig.~4 of \citet{Nut82} except that the data are different. The scatter of the data is rather large and the range of values regrettably rather small, but a linear dependency of $\ln n_1^2$ on $[T^3(\ln S)^2]^{-1}$ seems to exist. The least square fitting gave the following results for the coefficients of the linear relation for the corrected data
\begin{equation}
a=(4.40\pm0.61)\times10^{12}\;,\quad b=22.05\pm3.14\;.
\end{equation}  
From this, we obtain the following analytical approximation to the experimental data on the nucleation rate by SiO
\begin{equation}
J=n_1^2\exp\left((1.33\pm3.1)-{(4.40\pm0.61)\times10^{12}\over T^3(\ln S)^2}\right)
\label{NukSioExp}
\end{equation}
in units cm$^{-3}$\,s$^{-1}$. This may be used as a purely empirical fit formula to nucleation rates, the analytical form of which is motivated by theoretical considerations and the constants of which are determined from the results of laboratory experiment. 

\begin{figure}

\includegraphics[width=\hsize]{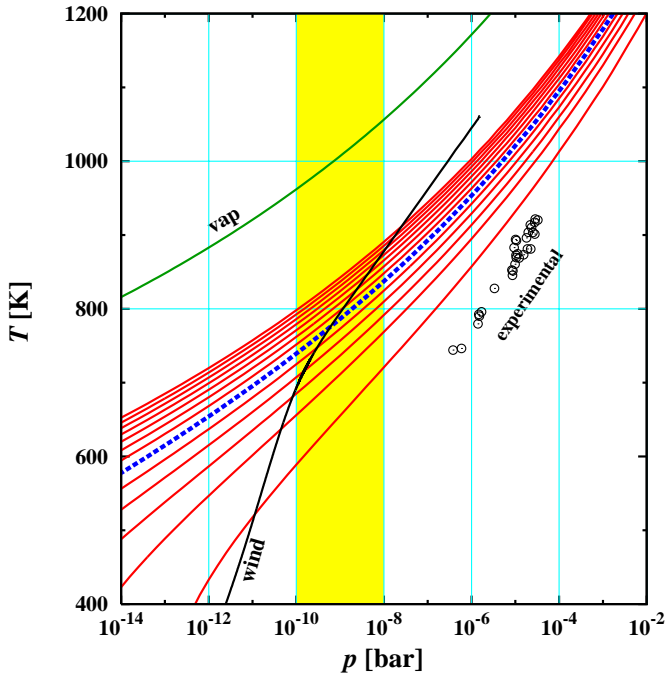}

\bigskip
\caption{Lines of constant nucleation rate per H-nucleus according to 
the numerical fit to the experimental data. The red lines correspond to
values of $J/N_{\rm H}$ from $10^{-34}$\,s$^{-1}$ (top) to $10^{-10}$ (bottom) in steps of~$10^2$. The shaded area indicates the range of pressures typically encountered at the inner edge of dust shells. The {\bf thick dashed (blue)} line corresponds to the critical nucleation rate of $J/N_{\rm H}=10^{-20}$ particles\,cm$^{-3}$\,s$^{-1}$ where under circumstellar conditions massive dust nucleation commences. The {\bf uppermost (green)} line corresponds to the stability limit if solid SiO in equilibrium with a gas with standard cosmic elemental abundances. The line denoted as ``wind'' corresponds to the $p$-$T$-trajectory of a wind model. The circles corresponds to the data points of the laboratory experiments, if the pressure of SiO in the condensation zone is converted into an equivalent pressure of a gas with cosmic elemental composition. 
}

\label{FigNcZel}
\end{figure}

\subsection{Circumstellar SiO-nucleation}

We assume that the nucleation rate calculated from Eq.~(\ref{NukSioExp}) remains applicable over a wider range of condensation conditions than that 
for which the experiments were performed because the coefficients in this expression are essentially determined by molecular and solid phase properties and not so much by the pressure-temperature conditions of the laboratory experiments. We use this to calculate nucleation rates based on clustering of SiO molecules for astrophysical problems.

Figure~\ref{FigNcZel} shows lines of constant nucleation rate in the $p$-$T$-plane per H-atom, $J'=J/N_{\mathrm{H}}$, calculated in this way for an oxygen rich element mixture. Here $N_{\mathrm{H}}$ is the particle density of H nuclei if all hydrogen atoms would be present as free atoms. The particle density of SiO molecules is assumed to be given by $n_1=\epsilon_{\mathrm{Si}}N_{\mathrm{H}}$, and the supersaturation ratio $S$ is calculated by using our fit-formula,
Eq. (\ref{VapPrSiO}), for the vapour pressure of SiO.

If we have a typical number of dust grains of the order of $3\times10^{-13}$ per hydrogen nucleus and assume that these are formed in about one year, then the formation rate would be about $10^{-20}$ seed particles per H nucleus per second. This is the black line in Fig.~\ref{FigNcZel}; it marks the critical pressure-temperature combination where massive dust particle formation occurs.
For typical pressures between $10^{-10}$ bar at mass-loss rates of $10^{-6}\,\rm
M_{\sun}\,a^{-1}$ and $10^{-8}$ bar at mass-loss rates of $10^{-4}\,\rm
M_{\sun}\,a^{-1}$ the critical rate corresponds to temperatures roughly between 720\,K and 850\,K. This range of gas temperatures for the onset of massive dust formation is predicted by the empirical nucleation rate (\ref{NukSioExp}).

For comparison, Fig.~\ref{FigNcZel} also shows the $p$-$T$ combinations that correspond to the data points of \citep{Nut82}. Their pressure of SiO in the condensation zone is converted to the total pressure in an element mixture with cosmic abundances via
\begin{equation}
p={p_1\over2\epsilon_{\rm Si}}
\end{equation}
withe Si abundance $\epsilon_{\rm Si}$ and neglecting He. This gives an impression on the extrapolation that is involved if the results are applied to circumstellar nucleation.

\begin{table*}
\caption{Models of spherically-symmetric and stationary dust-driven winds of AGB stars with different mass-loss lates, based on the empirical nucleation rate for SiO, equation (\ref{NukSioExp}).}

\begin{tabular}{lllllllllll}
\hline
\hline
\noalign{\smallskip}
Quantity & symbol & & & & & & & unit\\
\noalign{\smallskip}
\hline
\noalign{\smallskip}
& &\multicolumn{6}{c}{stellar parameters} & \\
Mass-loss & $\dot M$ &  1   & 2    & 3    & 5    & 10   & 20   & 30   & 50   & $10^{-5}\,\rm M_{\sun}yr^{-1}$ \\
\noalign{\smallskip}
Mass & $M_*$         & 1    & 1    & 1    & 1    & 1    & 1    &  1   &  1   & M$_{\sun}$ \\ 
Luminosity & $L_*$   & 5    & 5    & 5    & 5    & 5    & 10   & 10   &  10  & $10^3\,\rm L_{\sun}$ \\
Temperature & $T_*$  & 2700 & 2700 & 2700 & 2700 & 2700 & 2700 & 2700 & 2700 & K \\
Radius & $R_*$       & 323  & 323  & 323  & 323  & 323  & 456  &  456 & 456  & R$_{\sun}$ \\
\noalign{\smallskip}
& & \multicolumn{6}{c}{sonic point} & \\
\noalign{\smallskip}
Position & $r_{\rm c}$                  & 9.21  & 7.98  & 7.50  & 7.05  & 6.60  & 6.42  & 6.20  & 5.93 & $R_*$ \\
Temperature & $T(r_{\rm c})$            & 629.1 & 675.9 & 697.1 & 719.0 & 743.2 & 753.4 & 766.9 & 783.7 & K \\
Pressure & $p(r_{\rm c})$               & 1.75  & 4.85  & 8.35  & 16.0  & 37.1    & 39.5  & 64.2  & 118 & $10^{-10}$\,bar \\

Degree of condensation & $f(r_{\rm c})$ & 13.4  & 14.06 & 14.5  & 15.4  & 17.6  &  6.37 & 6.15  & 6.77 & \% \\
\noalign{\smallskip}
& & \multicolumn{6}{c}{dust shell} & \\
\noalign{\smallskip}
Terminal velocity & $v_\infty$          & 4.42 & 5.42   & 6.08 & 7.01  & 8.67  & 10.71 & 11.81 & 13.10  &  km\,s$^{-1}$ \\
Terminal condensation & $f_\infty$      & 19.4 & 26.9   & 33.9 & 47.2  & 71.4  & 62.7  & 75.8  & 88.3 & \% \\
Rosseland optical depth & $\tau_{\rm R}$          & 0.36 & 0.75   & 1.03 & 2.37  & 6.21  & 5.26  & 9.31  & 18.02 & \\
Visual optical depth & $\tau_{0.5\mu\rm{m}}$   & 0.97 & 2.01   & 3.26 & 6.39  & 16.75 & 14.19 & 25.10 & 48.60 & \\
\noalign{\smallskip}
\hline
\end{tabular}

\label{TabModls}
\end{table*}

\section{Models for dust-driven winds}

As a simple first application of the new nucleation rate some models for dust formation in outflows from M stars are calculated. The aim of this is to check which condensation temperatures of dust result if the nucleation rate for SiO is applied to circumstellar dust condensation. It is assumed that the condensation of silicate dust is triggered by the nucleation of SiO clusters from the gas phase and that later these seed particles grow to magnesium-iron silicate dust grains by collecting further SiO molecules and Mg and Fe atoms from the gas phase, while reaction with H$_2$O provides the required additional oxygen atoms for building the SiO$_4$ tetrahedrons of the silicate compounds.

In a stellar outflow more then one dust component may be formed in oxygen-rich matter \citep[e.g.][]{Gai98}. If some components are already formed at higher temperatures than the silicates (corundum for instance) and if their opacity suffices to drive the outflow velocities to supersonic values, then the gas density has strongly decreased by expansion if temperature drops below the stability limits of silicates. In that case silicate dust probably precipitates on top of the seed particles formed at higher temperatures and separate nucleation by SiO is probably not important. If, however, the subsonic-supersonic transition in the flow is due to silicate dust condensation, then the whole process is governed by nucleation of silicates. Then we have to deal with dust-driven winds. Here we perform some sample calculations for dust-driven, spherically symmetric and stationary wind models of giant stars; other possible cases where seed particles of a different origin are present will not be considered. 

\subsection{Model assumptions}

The wind models including dust formation and growth are calculated by the methods outlined in \citet{Gai84} and \citet{Gai85,Gai87}. In that papers, the condensation of graphite dust in C stars was considered, but the modifications required for calculating silicate dust condensation for M stars are simple and {\bf straightforward.} The models presented here are constructed exactly as described in \citet{Gai85}. We do not repeat, therefore, a description of the details of the method but only give remarks on a few points that change in case of M stars.

Since it is assumed that the condensation of silicate dust is triggered by nucleation of SiO clusters from the gas phase, the nucleation rate is assumed to be given by Eq.~(\ref{NukSioExp}). The final growth of macroscopic dust grains is assumed to result in the formation of amorphous silicates with an iron-rich composition. That such kind of material is formed in most circumstellar dust shells around M stars is known from analysis of their infrared spectra \citep[see][ for reviews]{Mol10,Hab96}. Additionally {\bf it has been known} since a few years from laboratory analysis of the composition of silicate grains from AGB stars \citep{Vol09a,Ngu10,Bos10} that amorphous silicate dust grains from AGB stars have widly varying iron contents, with iron-poor grains being rare. This is an important point because the extinction properties of the silicates in the near infrared and, hence, the possibility to drive a wind by radiation pressure on silicate dust strongly depends on the iron content of the silicates. It has not always adequately been considered in the literature that real amorphous silicate grains are \emph{not} iron-poor, except for the crystalline component. 

A compilation of the equations required for calculating the silicate dust and other dust components in M stars is given in \citet{Fer06}. Here we consider for simplicity only a single amorphous silicate dust component. Further we assume that this has the composition MgFeSiO$_4$, i.e., we do not include a calculation of the mixing ratio between the two end members of the solid solution in the present model calculation. A composition as the assumed one will be the outcome of non-equilibrium condensation in a rapidly cooling environment because of nearly equal abundances of Si, Mg, and Fe in the standard cosmic element mixture \citep[see, e.g.,][]{Lod09}. Such a composition is also compatible with the results of laboratory analysis of dust grains from AGB stars, though the observed scatter in composition is large \citep{Bos10}. 

The basic net chemical reaction for the silicate condensation is assumed to be
\begin{displaymath}
\rm SiO\ +\ Mg\ +\ Fe\ +\ 3H_2O\ \longrightarrow\ MgFeSiO_4(s)\ +\ 3H_2\,.
\end{displaymath}
For calculating the growth rate of dust grains \citep[see][]{Fer06} one needs the abundances of the gas phase species in this reaction equation. The composition of the gas phase is treated in a simple way. In oxygen-rich matter the Si in the gas  phase almost exlusive forms SiO and all O not bound in SiO and CO forms H$_2$O, while Mg and Fe are mainly present as free atoms. We simpy describe this as the gas phase composition and do not explicitely calculate its chemical composition.

The dust opacity is calculated with the complex index of refraction of the dirty silicate model of \citet[ their set 1]{Oss92} which is well suited to model the infrared emission from dust shells of M stars. The absorption and scattering efficiencies $C_\lambda$ are calculated in the small particle limit for spherical grains. From analysis of silicate dust grains from AGB stars identified im primitive meteorites one knows that the large majority of the grains have diameters significantly less than 1\,$\mu$m \citep{Vol09a,Ngu10,Bos10} which justifies {\bf the use of} this approximation from the red to the far IR spectral region, though at the shortward border of the wavelength region where absorption and scattering of stellar photons transmits momentum to the dust grains (mainly the range $0.5\lesssim\lambda\lesssim3\,\mu$m) one comes to the verge of validity of this approximation. Investigation of silicate grains from AGB stars \citep{Vol09a,Ngu10} show the grains to be somewhat elongated, though deviations from sphericity seem to be small for most particles. Therefore we assume for simplicity spherical grains in the calculation of dust opacities ({\bf in any case}, using a distribution of shapes, like CDE, does not significantly change the models).
 
For calculating the radiation pressure on the dusty gas one needs the flux average
\begin{equation}
\kappa_H=\int_0^\infty \kappa_\lambda^{\rm ext} H_\lambda\,{\rm d}\lambda\ \mbox{\Huge/}
\int_0^\infty H_\lambda\,{\rm d}\lambda\,.
\end{equation}
of the mass-extinction coefficient $\kappa_\lambda^{\rm ext}$ for the dust-gas mixture. Here $4\pi H_\nu$ is the spectral energy flux density of the radiation field. From radiative transfer calculations of circumstellar dust shells we found that this average can be well approximated by (for details see Appendix~\ref{Appendix1})
\begin{equation}
\kappa_H=\kappa_{\rm P}^{\rm ext}(T_*)\,{\rm e}^{-\tau_*}+\left[\kappa_{\rm P}^{\rm ext}(T_{\rm ph})+(1-f)\kappa_{\rm R}^{\rm ext}(T_{\rm d})\right]\left(1-{\rm e}^{-\tau_*}\right)
\label{ApproxKapH}
\end{equation}
where $\kappa_{\rm P}^{\rm ext}$ is the Planck-average and $\kappa_{\rm R}^{\rm ext}$ is the Rosseland-average of the extinction coefficient,  $T_*$ is the effective temperature of the star, $T_{\rm ph}$ the dust temperature at the inner edge of the dust shell, $T_{\rm d}$ the dust temperature in the shell, and $\tau_*$ is the optical depth calculated from the inner edge of the dust shell outwards using $\kappa_{\rm P}^{\rm ext}(T_*)$ as opacity. The Eddington factor $f$ can be taken, e.g., from the approxiation of \citet{Luc76} \citep[see {\bf also}][for details]{Gai87}.

\begin{figure}

\includegraphics[width=\hsize]{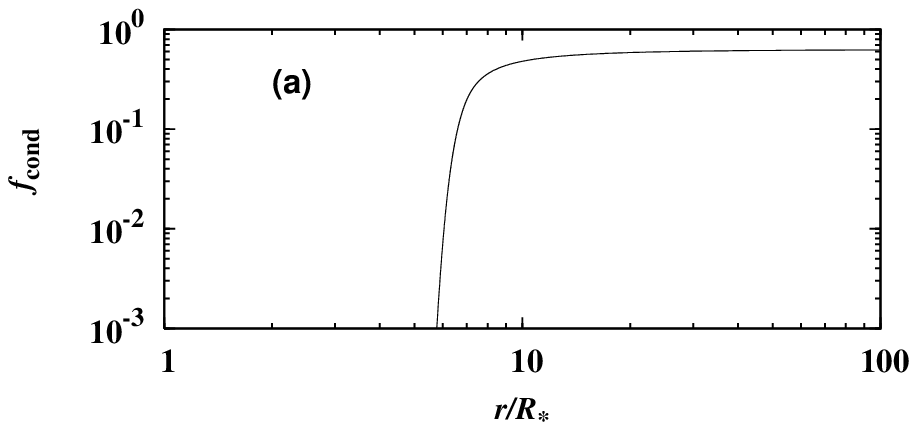}

\includegraphics[width=\hsize]{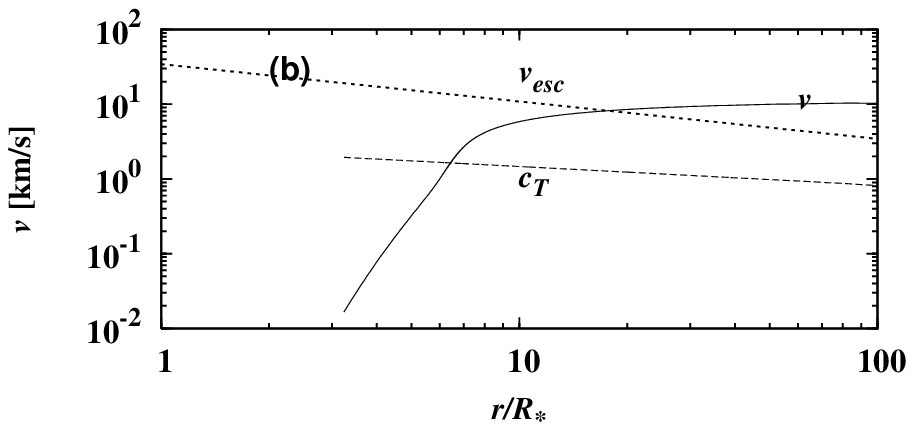}

\includegraphics[width=\hsize]{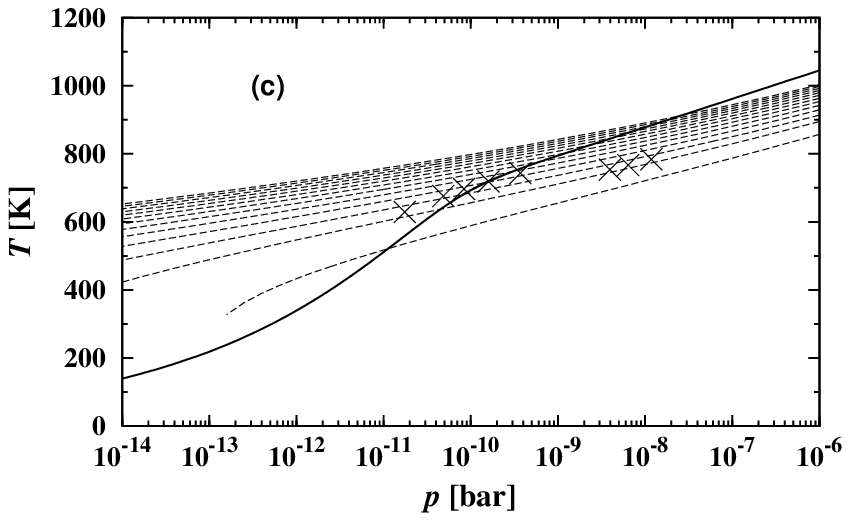}

\bigskip
\caption{Some results for a wind model with $\dot M=5\times10^{-6}\,\rm M_{\sun}yr^{-1}$. (a) Variation of the degree of condensation of Si into silicate dust with radial distance from the mass-centre. (b) Radial variation with distance of outflow velocity $v$ (solid line), sonic velocity $c_{\rm T}$ (dashed line), and escape velocity from star (dotted line). (c) Variation of $p$ and $T$ along the wind (solid line) and lines of constant nucleation rate (dashed lines). The crosses denote the pressure and temperature at the inner edge of the dust shell for the models of Table~\ref{TabModls}.}

\label{FigModl}
\end{figure}

\subsection{Model calculations}

With these assumptions some models are calculated for dust driven winds of oxygen-rich AGB starst. The assumption of stationarity can hardly be justfied for AGB stars because they are variables. The outerflowing gas is subject to shock waves triggered by the pulsation of the underlying star and the whole structure is far from being stationary. One can only speculate that stationary models describe long-term average properties of such outflows, which is justified, at least in part, by the success by which radiative transfer models based on $r^{-n}$ density distrbutions can be used to model IR spectra from dust shells. In any case, the purpose of our present calculation is to see if the new empirical nucleation rate results in condensation temperatures of silicate dust that principally agree with what one derives from analysis of IR spectra. 

Models are calculated for the higher end ($\dot M\gtrsim1\times10^{-6}\,\rm M_{\sun}yr^{-1}$) of the observed range of mass-loss rates for such stars ($\dot M\gtrsim1\times10^{-8}\,\rm M_{\sun}yr^{-1}$) because the present model does not consider drift of dust grains relative to the gas. This is not important for higher mass-loss rates because drift velocities are less than the thermal verlocity of the growth species in this case, such that the collision rate of dust grains with growth species is not significantly increased by drift. For low mass-loss rates, however, particle drift dramatically enhances dust growth and must be considered.

Table \ref{TabModls} shows the stellar parameters chosen for the calculations and some results. Figure \ref{FigModl} shows some details of the model with a mass-loss rate of $\dot M=5\times10^{-6}\,\rm M_{\sun}yr^{-1}$. Its general structure is representative for all our models. One feature of interest is the rather strong increase in the degree of condensation (Fig.~\ref{FigModl}a) which means that dust appears rather suddenly and the dust shell has a reasonably well-defined inner boundary. The wind trajectory shown in 
Fig.~\ref{FigModl}c shows how an outflowing gas parcel first cools and reduces pressure by expansion, then intrudes the region of increasing nucleation of seed particles for silicate dust growth, upon which the resulting strong acceleration results in a rapid blow-out of the gas parcel into low-pressure regions where finally growth ceases because of strong expansion of gas.

Of particular interest in our context are the typical temperatures at the onset of condensation, because this defines the hottest temperatures of dust grains that may contribute to the infrared spectrum from the dust shell. We take the position of the sonic point, where the wind velcity $v$ equals the sonic velocity $c_{\rm T}$ (see Fig.~\ref{FigModl}b) as the position of the inner boundary of the dust shell. The distance of this point is given in Table  \ref{TabModls} and also the temperature at this radius. At this point already substantial condensation has occured in order that radiation pressure on the dust component increases to the level that it locally surmounts the gravitational pull by the star. Additional substantial condensation, however, is requiered to enhance radiation pressure for ultimately lifting out the matter from the deep gravitational potential well of the star (see Fig.~\ref{FigModl}b). The degree of condensation at the sonic point and the final degree of condensation far in the outflow are shown in Table \ref{TabModls}. 

The table also shows the pressure and temperature at the inner boundary of the dust shell for the models. These data are also plotted in Fig.~\ref{FigModl}c as crosses. This outlines the typical range of ``condensation temperatures'' where avalanche nucleation and dust growth has just begun. The dust grains at the inner border of the dust shell are  the hottest dust grains existing for the considered dust species because of their closeness to the star. Their temperature may be derived from observations of infrared spectra and this temperature in turn can be compared with the theoretical results following from our approximation for the nucleation rate.

\section{Comparison with observations}

We do not present radiative transfer models for dust shells but use data from the literature for dust temperatures at the inner edges of dust shells. 

\begin{figure}

\includegraphics[width=\hsize]{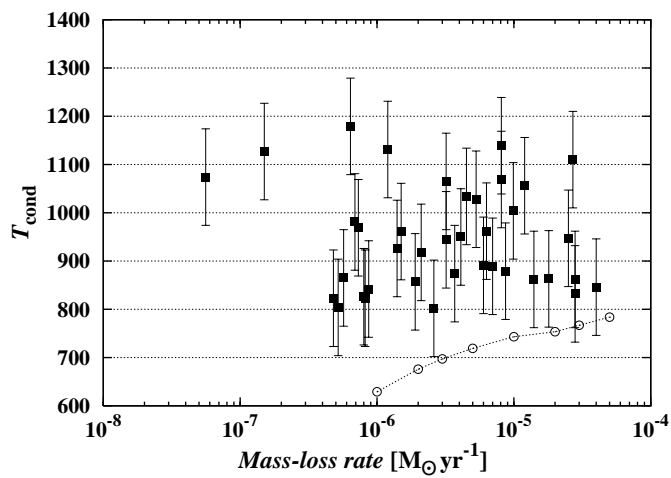}

\caption{Temperature at the inner edge of dust shells of AGB stars in the Small and Large Magellanic clouds with estimated errors of the temperature determination \citep[data from][]{Gro09}. This corresponds to the temperature where massive silicate dust condensation  commences in the outflow. The data are plotted against the derived mass-loss rate. Dotted line with open circles denotes the temperature at the sonic point of the models from Table \ref{TabModls}.}

\label{FigTcondObs}
\end{figure}

In such models for dust shells it is usually assumed that (i) the dust is formed at some condensation temperature $T_{\rm c}$ in the outflow and (ii) the density distribution of dust outwards from this point is proportional to $r^{-2}$. The mean intensity, $J_\nu$, is then calculated from a solution of the radiative transfer problem with the inner boundary condition that we have a star of given luminosity $L_*$ and effective temperature $T_*$. The dust opacity is calculated from optical constants of materials thought to be analogues of circumstellar dust materials. For M-stars different kinds of silicates and sometimes Al-Ca-bearing minerals are considered in the calculations. The run of temperature in the shell and its inner radius are determined self-consistently from radiative equilibrium between emission and absorption by the dust grains and the solution of the radiative transfer problem. 

The kind of dust material used for the modelling, the temperature $T_{\rm c}$, the corresponding inner radius $R_c$, and the dust density at $R_c$ then are varied to obtain an optimum fit with observed infrared spectra from circumstellar shells. If the outflow velocity is known from, e.g., the width of CO lines, one can determine from the dust density a dust mass-loss rate and with an assumed gas/dust ratio a total mass-loss rate.

It is generally assumed that the temperature $T_{\rm c}$ derived this way reflects the dust temperature at the inner edge of the dust shell, where efficient dust formation commences. For this reason, this temperature is called the \emph{condensation temperature} of the dust. It is to be observed, that this temperature corresponds to the lattice temperature in radiative equilibrium of particles, that have grown to such a size, that their extinction properties have changed to solid-like behaviour. This does not imply any information on the gas temperature in that region.

From the numerous such studies published in the literature we consider here a set of data for late-type M-stars on the AGB from the Small and Large Magellanic clouds derived by \citet{Gro09} because this is a very homogeneous data set both with respect to the acquisition of spectra and the model data derived for all stars with the same method and the same set of opacity data. The condensation temperatures $T_{\rm c}$ are shown in Fig.~\ref{FigTcondObs}. Here only data for those objects are used for which the dust temperature at the inner edge could uniquely be determined by the model fit. The figure suggests that there exist two different groups of objects with different condensation temperatures.

For one group, the typical temperatures are around 880\,K$\pm100$\,K. These are the ``condensation temperatures'' of silicate dust that have to be explained by theoretical calculations of dust formation. 

Then there seems to exist a second group of objects with condensation temperatures between 1\,100\,K and 1\,200\,K. They could correspond to objects with observable production of Ca-Al-bearing dust species which are thermally more stable than silicates. We neglect such objects because they may represent physically distinct objects. 

Figure \ref{FigTcondObs} shows the temperature at the inner inner edge of dust shells in our theoretical models, based on the approximation for the nucleation rate from Sect.~\ref{SectNuc} as dotted line with circles. At first glance the theoretical condensation temperatures seem to fall short of what is followed from radiative transfer models for observed dust shells by about 100 to 200\,K. This cannot result from the big uncertainties connected with the conversion of the experimental data of \citet{Nut82} into the apprimate formula (\ref{NukSioExp}) because the nucleatio n rate depends very steeply on temperature, as can be seen in Fig.~\ref{FigNcZel}. Variations in $J$ over even many decades correspond to relativ small temperature changes. The discepancy is too big to result from such errors.

\begin{figure}[t]

\includegraphics[height=\hsize,angle=-90]{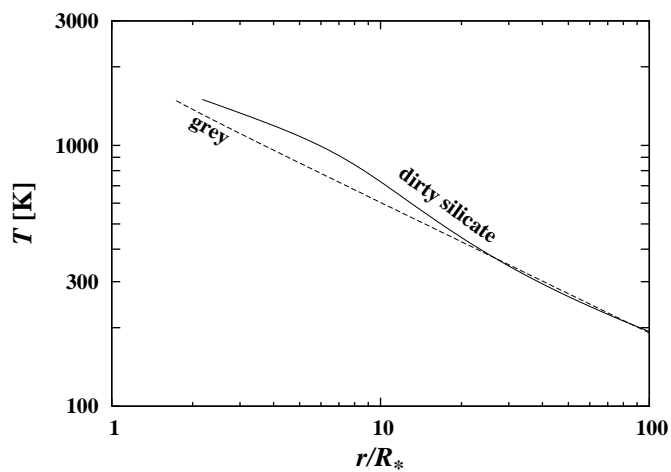}

\caption{Dust temperature and grey temperature in optically thin dust shells for a star radiating like a black body with temperature $T_{\rm eff}=2\,700$\,K, calculated for silicate dust with opacities from \citet{Oss92}.
}

\label{FigTempDu}
\end{figure}

A more likely origin of the difference in condensation temperature is the approximation that the model does not discriminate between gas temperature, that rules nucleation, and dust temperature, that rules the emission of IR radiation by dust. Generally both temperatures are significantly different, in particular if dust shells are optically thin ($\tau_{\rm R}\lesssim 1$). For optically thin to moderately optically thick dust shells the local mean radiation intensity is dominated by the stellar contribution. The dust temperature in radiative equilibrium is then given by the well known relation
\begin{equation}
T_{\rm d}=W^{1\over4}T_*\left({\kappa_{\rm P}(T_{\rm eff})\over\kappa_{\rm P}(T_{\rm d})}\right)^{1\over4}>W^{1\over4}T_*=T_{\rm grey}\,,
\label{EqTempRadEq}
\end{equation}
with the geometrical diluton factor
\begin{equation}
\quad W=\frac12\left(1-\sqrt{1-{R_*^2\over r^2}}\right)\,.
\end{equation}
Here $R_*$ the stellar radius, $r$ the radial distance to the stellar centre, $T_*$ the effective temperature of the star and $T_{\rm grey}$ the temperature of a grey absorber. The variation of the Planck-average
\begin{equation}
\kappa_{\rm P}(T)=\int_0^\infty \kappa_\nu B_\nu(T)\,{\rm d}\nu\ \mbox{\LARGE/}
\!\int_0^\infty B_\nu(T)\,{\rm d}\nu\,.
\end{equation}
of the dust absorption coefficient with temperature has a significant influence on the dust temperature.

Figure \ref{FigTempDu} shows the dust temperature of silicate dust material in radiative equilibrium at different distances from a central star with effective temperature $T_{\rm eff}=2\,700$\,K, determined  according to Eq.~(\ref{EqTempRadEq}), using optical data from \citet{Oss92}. Additionally  the temperature of a material with grey absorption is shown, for which the monochromatic absorption is independent of wavelength. For silicates the dust temperature is higher than for a grey body because their Planck-averaged absorption coefficient is \emph{higher} for the effective temperature of the stellar radiation field than for the much lower dust temperatures. Since the gas temperature tends to be close to the grey temperature except behind shocks \citep[cf.][]{Sch03}, this behaviour can easely explain the remaing difference between the condensation temperature in our models and the observed condensation temperatures. Hence, the laboratory determined nucleation rate for SiO \citep{Nut82}, corrected for the the strong reduction of vapour pressure of SiO required by recent measurements \citep{Wet12,Nut06} seems to be able to explain the dust condensation in circumstellar dust shells of AGB stars.

Unfortunately we presently cannot definitely fix the difference between gas and dust temperature unless the model calculation considers exlicitely a different gas and dust temperature. Such more elaborate model calculations will be published elsewhere.

\section{Concluding remarks}

We have re-analyzed the long-standing problem of nucleation of silicate dust grains in circumstellar dust shells around oxygen-rich evolved stars. Though it has been proposed already at an early stage that this is due to clustering of the abundant SiO molecules \citep{Don81,Nut82}, there remained serious problems with this hypothesis. The nucleation process was studied by dedicated laboratory expriments by \citet{Nut82,Nut83}, but the results, if applied to ciurcumstellar conditions, seemed to predict much too low condensation temperatures \citep{Gai86} compared to the dust temperatures at the inner edge of dust shells derived from infrared spectra.  

New determinations of the vapor pressure of SiO by \cite{Nut06,Fer08} and that reported on in this paper \citep[see also][]{Wet12} show that former determinations were grossly in error. Our new measurements yield vapour pressures even somewhat lower than that obtained by \citep{Fer08}.

\citet{Nut06} showed already that the case for SiO nucleation may be not as unfavourable as was thought in the past. By using their vapour pressure data to calculate more realistic supersaturation values $S$ and to determine from this and their approximation to the nucleation rate $J$ in \citet{Nut82} they showed that condensation temperatures were much increased. But still there remains a significant gap to observed values which they try to explain by non-LTE effects in population of energy levels in SiO \citep[cf. also][]{Nut81}.

Here we performed a complete re-evaluation of the experimatally determined data for SiO nucleation of \citet{Nut82} with our new vapour pressure data and derived a fit-formula for the nucleation rate. This can be used in model calculations for stellar outflows. Some explorative calculations were done by calculating stationary dust-driven wind models in a simple approximation. The resulting temperatures at the inner edge of the dust shells were compared to data for the dust condensation temperature derived from analysis of IR spectra. It was found that model results based on the new approximation for the nucleation rate are now close to what one observes. We argue, that the remaining small gap is due to the difference between gas temperature (responsible for nucleation) and dust temperature (responsible for IR emission) that was not considered in our simple models.

Our result suggests that, indeed, in circumstellar dust shells around M stars silicate dust formation may well start with nucleation of SiO. 

This holds only for such cases where not some different kind of seed particles, formed well before SiO nucleation becomes possible (e.g. corundum grains), served as growth centres for the silicate grains. Observed condensation temperatures suggest, that both cases seem to occur in circumstellar shells.

The problem of silicate nucleation can, however not be considered as completely solved, because our model calculations assumed stationary outflows, which are at best a rather crude approximation for winds of AGB stars, because all such stars are variables. Time dependent model calculations accounting for the shock structure in the winds and accounting for the difference between gas and dust temperature are required, before definite conclusions can be draw.
  

\begin{acknowledgements}
This work was supported in part by Forschergruppe 759' and special research progamm SPP 1385  which both are supported by the `Deutsche Forschungs\-gemeinschaft (DFG)'. 
\end{acknowledgements}

\begin{appendix}
\section{Approximation for flux-averaged extinction}
\label{Appendix1}

In a study of radiative transfer models in circumstellar dust shells (Gail et al., in preparation) it was found that the flux-averaged extinction coefficient can be approximated rather well by the expression given in Eq.~(\ref{ApproxKapH}). This approximation is based on the following considerations:

1. At the inner edge of the dust shell the radiation field is dominated by the stellar radiation. The spectral distribution of the flux, $4\pi H_\lambda$, can be approximated by a black body with the effective temperature of the star. The flux averaged extinction coefficient can be approximated by the Planck-averaged extinction coefficient $\kappa_{\rm P}^{\rm ext}(T_*)$ in this case.

2. Going from the inside into the shell, the stellar contribution to the radiation field diminishes as $\exp(-\tau_*)$, where $\tau_*$ ist the optical depth calculated from the inner edge with $\kappa_{\rm P}^{\rm ext}(T_*)$. At the same time, the contribution of the local dust emission to the total flux increases proportional to $1-\exp(-\tau_*)$.

3. As long as the optical depth of the shell is not very high, the spectral flux distribution from dust emission can be approximated by a black body radiation field with temperature corresponding to the inner edge of the dust shell, $T_{\rm ph}$. This results in a contribution $\kappa_{\rm P}^{\rm ext}(T_{\rm ph})$ to the flux averaged extinction coefficient.

4. If the shell becomes optically very thick, the flux-averaged extinction coefficient should approach the Rosseland mean of the local radiation field, $\kappa_{\rm R}^{\rm ext}(T_{\rm d})$. The approach to the optically thick case can be modelled by the factor $1-f$ with the Eddington factor $f$, which is $f\approx1$ in an optically thin shell and approaches $f\approx1/3$ for optically thick shells.

\begin{figure}[t]

\centerline{
\includegraphics[width=\hsize]{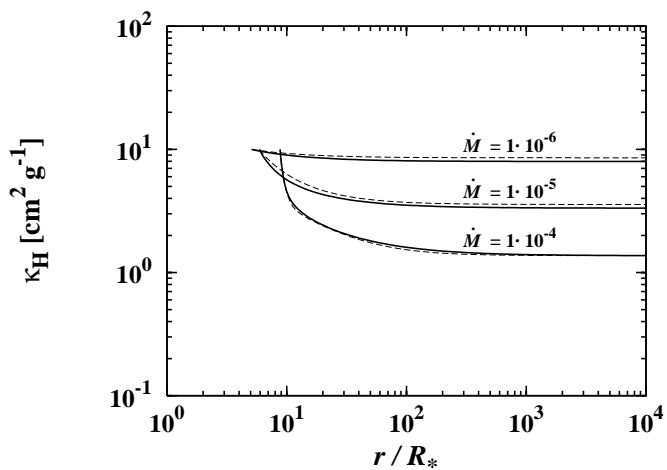}
}

\caption{Approximation of the flux-averaged extinction coefficient 
$\kappa_{H}$ in models for a dust shell with a powerlaw wavelength variation of the extinction coefficient for different mass-loss rates (in units M$_\odot$\,a$^{-1}$). Solid line: Result of a complete model calculation of radiative transfer. Dashed line: Approximation according to Eq. (\ref{ApproxKapH}).
}

\label{FigApproxKapH}
\end{figure}

As a demonstration for the accuracy of the approximation Fig.~\ref{FigApproxKapH} compares the result for $\kappa_{H}$ based on a full calculation of radiative transfer, as outlined in \citet{Wet12a}, using a powerlaw variation $\kappa_\lambda\propto\lambda^{-1}$ as a simple model case. An inspection of the figure shows that the approximation for $\kappa_{H}$ is obviously rather accurate.

\end{appendix}


\end{document}